# Texture- and Stress-Dependent Electromechanical Response in Ferroelectric PZT: Insights from a Micromechanical Model

Saujatya Mandal, Debashish Das[*]

Department of Mechanical Engineering, Indian Institute of Science, Bengaluru, Karnataka 560012, India

## Abstract

The electromechanical response of $PbZr_{0.52}Ti_{0.48}O_3$ (PZT) near the morphotropic phase boundary (MPB) is strongly influenced by crystallographic texture and residual stress, both of which affect domain switching behavior. While these effects are critical for optimizing sensors, actuators, and MEMS devices, their combined influence remains poorly understood. We present a computational micromechanical model that captures texture- and stress-dependent polarization switching in MPB PZT. The framework incorporates both tetragonal and rhombohedral domain switching, along with interphase transformations, enabling accurate simulation of nonlinear electromechanical behavior. The model reproduces key experimental trends, including enhanced piezoelectric response in (001)-textured ceramics, and degradation under high in-plane stress. The implementation, provided as open-source MATLAB code, offers an accessible platform for experimentalists and materials designers to explore and interpret electromechanical behavior. By linking microstructural orientation and stress state to macroscopic response, this work provides a practical tool for understanding and designing next-generation piezoelectric materials.

*Keywords:* Morphotropic phase boundary (MPB) PZT, Crystallographic texture and residual stress, Ferroelectric domain switching, Electromechanical response modeling, Open-source MATLAB code

## 1. Introduction

Ferroelectric materials, particularly lead zirconate titanate ($PbZr_{0.52}Ti_{0.48}O_3$, PZT), are widely used in sensors, actuators, energy harvesters, RF switches, inkjet printer heads, and Micro-Electromechanical Systems (MEMS) due to their exceptional piezoelectric properties [1–6]. These properties arise from PZT's perovskite crystal structure and its morphotropic phase boundary (MPB), where rhombohedral and tetragonal phases coexist [7]. The MPB enhances electromechanical coupling by allowing multiple energetically equivalent polarization states, promoting domain switching and mobility.

---

[*] Corresponding Author: ddas@iisc.ac.in

However, the macroscopic performance of PZT is highly sensitive to crystallographic texture and residual stress, which strongly influence domain switching, ferroelasticity, and piezoelectric performance [8–12]. In bulk ceramics, grain texturing narrows the performance gap with single crystals while maintaining scalable processing routes [8,13–15]. Several fabrication techniques—including magnetic field alignment, hot press sintering, and templated grain growth—have enabled highly textured ceramics with enhanced piezoelectric properties [16–18]. For example, <001>-textured PZT produced by tape casting has demonstrated nearly double the $d_{33}$ of conventional ceramics [8], while other approaches have achieved up to 60% improvement in field-induced strain [13]. In thin-film PZT, both crystallographic orientation and residual stress play an even more dominant role. Orientation control via deposition techniques such as chemical solution deposition (and sol-gel processing), RF magnetron sputtering, metal-organic chemical vapor deposition (MOCVD), and pulsed laser deposition (PLD) has led to significant gains in performance. Films with strong (001) orientation have exhibited a ~2.5× improvement in effective piezoelectric response compared to (111)-textured counterparts [10,15], consistent with studies reporting higher remanent polarization and piezoelectric coefficients in highly (001)-oriented PZT thin film [19].

Although most studies report improved ferroelectric response for (001)/(100) textures, others have shown improved responses in {111} [20], {110} [21], or even randomly oriented [22] MPB PZT. These inconsistencies often arise from differences in residual stress, which is difficult to isolate experimentally. Thin films deposited on rigid substrates, such as platinized silicon, experience biaxial tensile stress that restricts domain wall motion and reduces field-induced strain [12,23]. Moreover, stress levels can vary with texture [24] and deposition method, further complicating interpretation [11,25]. A comprehensive understanding of texture–stress interactions is therefore critical, especially at the MPB, where switching mechanisms are inherently complex. However, experimentally isolating the individual roles of texture and stress remains challenging—particularly near the MPB—highlighting the need for complementary modeling approaches.

To systematically explore the role of these coupled variables, computational modeling offers a valuable complementary tool. Computational models can help decouple these intertwined effects, but most existing approaches focus on simpler tetragonal ferroelectrics [26–28]. Rhombohedral materials are less frequently modeled due to their greater number of polarization variants [29,30]. Only a few models consider MPB compositions explicitly [31,32], and even fewer incorporate crystallographic texture or residual stress. Notable efforts include self-consistent crystal plasticity models [33] and hybrid Finite Element Modeling (FEM)- virtual Piezoresponse Force Microscopy (PFM) approaches [34,35] that show how grain orientation, local stress, and epitaxial strain influence switching. However, these methods either focus on single-phase tetragonal PZT neglecting the complexities of MPB compositions, or are computationally expensive, and do not readily capture key experimental observables, such as butterfly loops and polarization–electric field hysteresis.

In contrast to previous models that focus solely on tetragonal switching or require high computational cost, this work develops a micromechanical model—based on the framework of Hwang *et al.* [26]—that explicitly incorporates both tetragonal and rhombohedral domain switching, including interphase transformations, to simulate polarization behavior near the MPB. The model reproduces key experimental trends such as butterfly loops and polarization–electric field hysteresis, while capturing the effects of crystallographic texture and residual stress. Provided as open-source MATLAB code (see **Supplementary Material**), the framework offers an accessible yet physically grounded tool for experimentalists and theorists alike to explore and design high-performance ferroelectric materials.

## 2. Theoretical Model and Implementation

We extend the micromechanical framework developed by Hwang et al. [26] to explicitly include both tetragonal and rhombohedral domain switching, as well as interphase transformations at the MPB. A detailed mathematical description and derivations of key equations are provided in the **Supplementary Material**.

We briefly review the main equations here. Crystallites are modeled as spherical inclusions in a homogeneous non-transforming polycrystalline matrix subjected to external applied electric field $E_i^\infty$ and stress $\sigma_{ij}^\infty$. The inclusion is constrained by the surrounding matrix, which already possesses a remanent strain $\varepsilon_{ij}^{rm}$ and remanent polarization $P_i^{rm}$, reflecting prior switching events. The inclusion undergoes a transformation strain $\Delta\varepsilon_{ij}^t$ and polarization change $\Delta P_i^t$, leading to an energy change, $\Delta U$, given by [26]:

$$\Delta U = V_I\left[-\sigma_{ij}^\infty \Delta\varepsilon_{ij}^t - E_i^\infty \Delta P_i^t + \frac{(7-5\nu)Y}{15(1-\nu^2)}\left(\varepsilon_{ij}^s - \varepsilon_{ij}^{rm} + \frac{1}{2}\Delta\varepsilon_{ij}^t\right)\Delta\varepsilon_{ij}^t + \frac{1}{3\kappa}\left(P_i^s - P_i^{rm} + \frac{1}{2}\Delta P_i^t\right)\Delta P_i^t\right] \quad (1)$$

where $P^s$, $\varepsilon^s$ are the spontaneous polarization vector and spontaneous strain tensor, respectively of each crystallite, the magnitude of which depends on tetragonal or rhombohedral symmetry. $Y$, $\nu$, and $\kappa$ are the Young's modulus, the Poisson's ratio, and the dielectric permittivity of the crystallite. A simplified representation of dielectric and elastic tensors (assumed same whether tetragonal or rhombohedral) is adopted to reduce computational cost while capturing essential physics. However, the code can be modified to account for changes in material properties. $V_I$ represents the volume of the spherical inclusion.

The first two terms represent the driving forces for switching, corresponding to the mechanical work (stress-induced energy contribution) and electrical work (electric field-induced polarization change). The last two terms act as penalty terms, incorporating the mechanical mismatch energy and the electrical (depolarization-field) mismatch energy. The mechanical mismatch penalty accounts for elastic energy costs during switching, while the electrical mismatch penalty reflects the interfacial energy barrier due to polarization discontinuities between domains of different phases. These penalties impose energetic constraints that resist spontaneous switching, ensuring stability of the ferroelectric phase under applied mechanical and electrical loads.

Domain switching is determined by thermodynamic energy minimization: switching occurs when the potential energy reduction exceeds a critical threshold $(-\Delta U \geq (V_I \Delta\Psi_c))$, representing the energy barrier for switching. From Eqn. 1, the switching criterion becomes [26,36]:

$$\sigma_{ij}^\infty \Delta\varepsilon_{ij}^t + E_i^\infty \Delta P_i^t - \frac{(7-5\nu)Y}{15(1-\nu^2)}\left(\varepsilon_{ij}^s - \varepsilon_{ij}^{rm} + \frac{1}{2}\Delta\varepsilon_{ij}^t\right)\Delta\varepsilon_{ij}^t - \frac{1}{3\kappa}\left(P_i^s - P_i^{rm} + \frac{1}{2}\Delta P_i^t\right)\Delta P_i^t \geq \Delta\Psi_c \quad (2a)$$

$$\alpha\left[\sigma_{ij}^\infty - \frac{2}{5}\bar{Y}\left(\varepsilon_{ij}^s - \varepsilon_{ij}^{rm} + \frac{1}{2}\Delta\varepsilon_{ij}^t\right)\right]\Delta\varepsilon_{ij}^t + \left[E_i^\infty - \frac{1}{3\bar{\kappa}}\left(P_i^s - P_i^{rm} + \frac{1}{2}\Delta P_i^t\right)\right]\Delta P_i^t - \Delta\Psi_c \geq 0 \quad (2b)$$

where $\alpha$ is a weighting factor that distinguishes between mechanically driven and electrically driven switching. The terms $\bar{Y}$ and $\bar{\kappa}$ (Eqn. 2b) represent an effective modulus and effective dielectric permittivity which are essentially fitting parameters. When Poisson's ratio is approximated as 1/2, a numerical factor of 2/5 is obtained in Equation (2b).

The third-order piezoelectric strain tensor, $d_{ijk}$, is simplified from three to two indices for the sake of notation, though the model maintains the full three-dimensional tensor representation. In the phenomenological thermodynamic model of Haun *et al.* [37], the piezoelectric coefficients are derived from electrostrictive relations rather than directly imposed symmetry constraints. For tetragonal and rhombohedral PZT, the derived piezoelectric coefficients are expressed in matrix form in Eqn. 3. The coefficients in Table 1 serve as input parameters for computing the effective piezoelectric response as a function of crystallographic orientation and phase content. Unlike tetragonal PZT, where $d_{33}$ is maximized along [001] (the spontaneous polarization direction), the maximum effective $d_{33}$ in rhombohedral PZT occurs at an angle ~56.7° from [111], rather than along the spontaneous [111] polarization direction.

$$d_{mn}^{tetragonal} = \begin{bmatrix} 0 & 0 & 0 & 0 & d_{15} & 0 \\ 0 & 0 & 0 & d_{15} & 0 & 0 \\ d_{31} & d_{31} & d_{33} & 0 & 0 & 0 \end{bmatrix} \quad (3a)$$

$$d_{mn}^{rhombohedral} = \begin{bmatrix} d_{33} & d_{31} & d_{31} & d_{14} & d_{15} & d_{15} \\ d_{31} & d_{33} & d_{31} & d_{15} & d_{14} & d_{15} \\ d_{31} & d_{31} & d_{33} & d_{15} & d_{15} & d_{14} \end{bmatrix} \quad (3b)$$

**Table 1**: Theoretical piezoelectric coefficients as a function of composition at 25°C [37]

| **Tetragonal (40/60) [40% PbZrO₃ & 60% PbTiO₃] (pC/N or pm/V)** | **Rhombohedral (60/40) [60% PbZrO₃ & 40% PbTiO₃] (pC/N or pm/V)** | **MPB (50/50) [50% PbZrO₃ & 50% PbTiO₃] (pC/N or pm/V)** | |
|---|---|---|---|
| | | Tetragonal | Rhombohedral |
| $d_{33}$ = 162 | $d_{33}$ = 189 | $d_{33}$ = 327 | $d_{33}$ = 624 |
| $d_{15}$ = 169 | $d_{15}$ = 60 | $d_{15}$ = 624 | $d_{15}$ = 150 |
| $d_{31}$ = −58.5 | $d_{14}$ = −25.1 | $d_{31}$ = −156 | $d_{14}$ = −110 |
| | $d_{31}$ = −80.5 | | $d_{31}$ = −310 |

The transformation of piezoelectric strain tensor $d_{ijk}$ follows the standard tensor transformation rule: $d'_{pqr} = A_{pi}A_{qj}A_{rk}d'_{ijk}$, where $A$ is the second-order coordinate transformation tensor that maps from the initial to the new coordinate system. The $d_{mn}$ components given in Table 1 are expressed relative to a cubic coordinate system with Z-axis ∥ [001], for both tetragonal and rhombohedral symmetry. In contrast, in Du et al.'s work [38], for tetragonal symmetry Z-axis ∥ [001], whereas for rhombohedral symmetry Z-axis ∥ [111]. It is important to note that despite the inherent lattice distortions in the rhombohedral system, the same transformation equation is applied under the assumption that these deviations are small enough to justify treating the transformation as if it were Cartesian, simplifying the analysis while still capturing essential trends in electromechanical behavior.

Figs. 1(a, b) illustrate the variation of $d_{33}$ for tetragonal (40/60) and rhombohedral (60/40) PZT crystals. In the tetragonal phase, $d_{33}$ reaches its maximum value of 162 pC/N along the polarization direction [001], while in the [111] direction, it decreases to 74 pC/N. The behavior of $d_{33}$ in the rhombohedral phase exhibits a different trend. Instead of peaking along the polarization direction [111], the maximum occurs approximately 56.7° away from it, a direction equivalent to [001] in perovskites. For rhombohedral (60/40) crystals, $d_{33}$ attains a maximum of 189 pC/N along the three [001] directions, while along the [111] directions, it is significantly lower at 71 pC/N. At the morphotropic phase boundary (MPB), the piezoelectric response is further enhanced, Figs. 1(c, d) with tetragonal crystals exhibiting a peak $d_{33}$ value of 327 pC/N along [001], whereas for rhombohedral crystals, the maximum value increases to 624 pC/N along [001].

The phenomenological model of piezoelectric coefficients successfully captures the enhanced electromechanical properties near the MPB [39], making it useful for optimizing texture-dependent piezoelectric performance in this study.

The model employs an incremental loading scheme where an external field—electrical, mechanical, or a combination—is applied in small steps, ensuring that only a limited number of crystallites switch at each stage. After each load increment, every crystallite is evaluated against the switching criterion, Eqn. 2, to determine whether a phase transformation occurs. Fig. 2 presents the numerical procedure in a flowchart. The simulation begins by defining the number of crystallites, phase fractions, and incremental loading steps. Using MATLAB's random number generator, Euler angles are assigned to each crystallite, ensuring a particular texture in the material and a representative mixture of tetragonal and rhombohedral phases characteristic of the MPB.

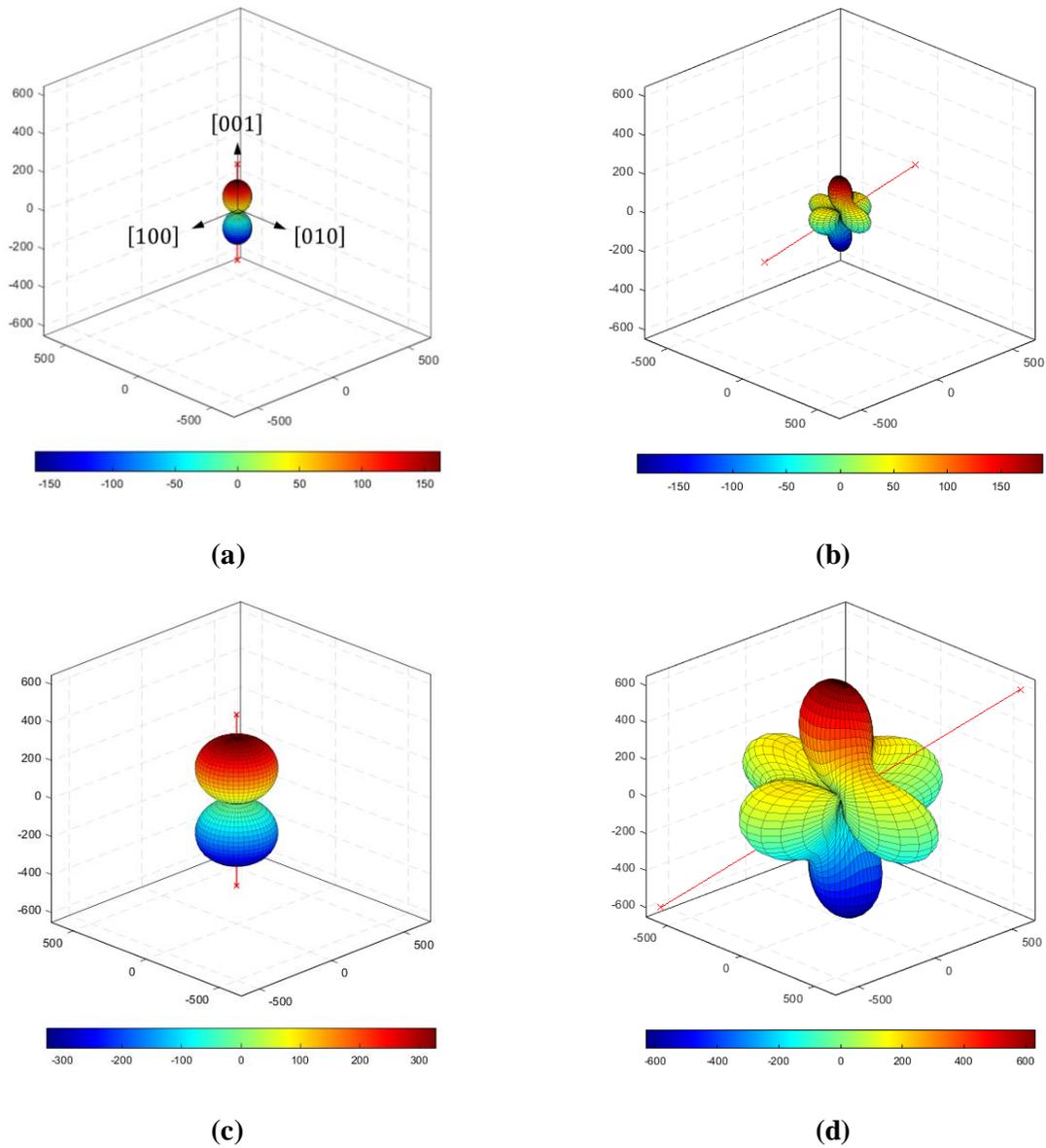

**Figure 1.** Visualization of $d_{33}$ for different PZT compositions: **(a)** Tetragonal (40/60), **(b)** Rhombohedral (60/40), **(c)** Tetragonal at MPB (50/50), and **(d)** Rhombohedral at MPB (50/50). The cubic axes in (a) are consistent across all subfigures. Polarization directions are indicated by red lines with end-crosses.

As the load is gradually increased, the system allows incremental switching of crystallites, mirroring real material behavior. The switching process follows energy minimization principles, evaluating all potential switching pathways and selecting the one that maximizes the driving force.

The coercive field for interphase switching is typically higher than that for intra-phase switching. As a result, intra-phase domain switching occurs first under low-field conditions. However, once the coercive field for interphase switching is exceeded, phase transformation contributes significantly to the enhanced electromechanical response.

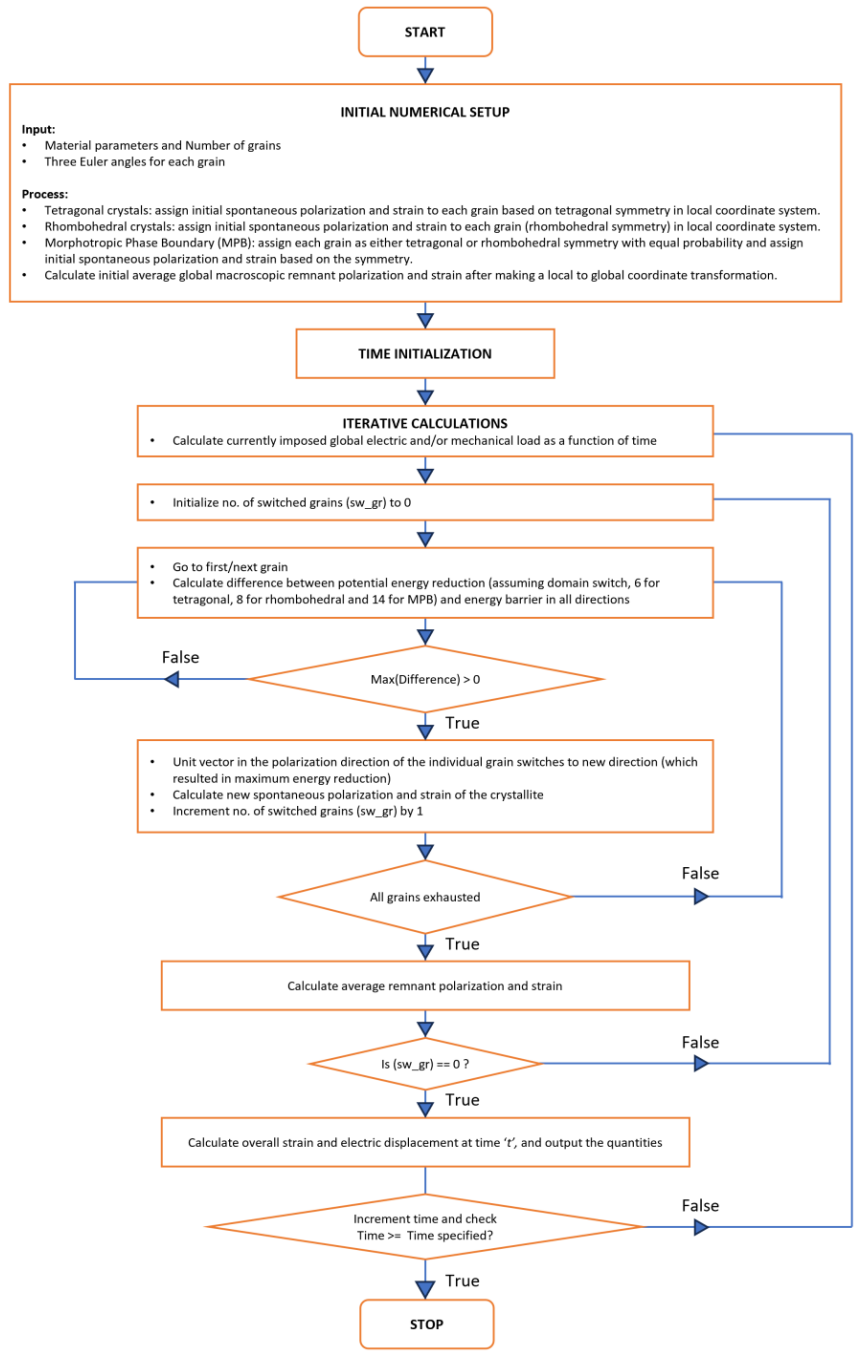

**Figure 2.** Flowchart summarizing the incremental loading and domain switching algorithm used in the numerical implementation.

Once all permissible switches are executed, the macroscopic remanent strain ($\varepsilon^{rm}$) and remanent polarization ($P^{rm}$) are computed by averaging over all crystallites. These values are updated iteratively, permitting additional switching without further load increments, until no further switching occurs.

The linear contributions to strain and electric displacement are computed for each crystallite using the elastic compliance, dielectric permittivity, and piezoelectric strain tensors, incorporating phase-specific components for tetragonal and rhombohedral symmetries. The macroscopic electric displacement and strain are then obtained as a volume average across all crystallites, capturing the enhanced electromechanical coupling resulting from the interaction of tetragonal and rhombohedral phases. This approach offers both physical insight and a computationally efficient tool for predicting texture- and stress-dependent responses in MPB ferroelectrics, facilitating comparisons with experimental data and guiding materials design. This micromechanical framework thus forms a robust foundation for interpreting experimental trends and optimizing ferroelectric device performance via texture and stress control.

## 3. Results and Discussion

### 3.1. Fitting Parameters

The fitting parameters used in this study, as detailed in Table 2, were sourced from Refs. [26,32,36], where they were derived by aligning numerical simulations with experimental data for polycrystalline ceramics. Several key considerations guided the parameter selection. The coercive field for intraphase 71° and 109° switching was set to half that of intraphase 90° switching, consistent with experimental observations in Ref. [32]. Additionally, the spontaneous polarization ratio between rhombohedral and tetragonal phases was maintained as reported in Ref. [32] to ensure consistency with prior studies. The piezoelectric coefficients ($d_{ijk}$) were scaled to ten times the values listed in Table 1, as Hwang *et al.* [26] used values an order of magnitude higher in their study. The primary focus of this work is not on parameter optimization but rather on understanding the mechanics of domain switching and elucidating the role of texture in shaping the electromechanical response. Therefore, a direct fitting of the model to PZT experimental data has been conducted at the end of this section to validate the numerical predictions.

**Table 2.** Parameters used in numerical simulation.

| Parameter | Description | Value [Reference] |
|---|---|---|
| $\bar{E}_0^{90}$ | Effective coercive field for intraphase 90° switch | 0.13 MV/m [26] |
| $\bar{E}_0^{180}$ | Effective coercive field for intraphase 180° switch | 1.0 MV/m [26] |
| $\bar{E}_0^{71\_109}$ | Effective coercive field for intraphase 71° and 109° switch | 0.064 MV/m [32] |
| $\bar{E}_0^{55\_125}$ | Effective coercive field for interphase 55° and 125° switch | 0.5 MV/m |
| $\bar{\kappa}$ | Effective dielectric permittivity | 0.80 µF/m [26] |
| $\bar{Y}$ | Effective elastic modulus | 7.5 GPa [26] |
| $P_0$ | Spontaneous polarization | 0.3 C/m² [26] |
| $\varepsilon_0^T$ | Spontaneous strain (tetragonal) | 0.0028 [26] |
| $\varepsilon_0^R$ | Spontaneous strain (rhombohedral) | 0.0023 [32] |
| $\alpha$ | Weight Parameter | 1.4 [26] |
| $d_{ijk}$ | Piezoelectric strain coefficient | [Table 1 × 10] [26] |
| $N$ | Number of Crystallites | 5000 [26] |

## 3.2. Validation of Numerical Scheme

To verify the correctness of the numerical implementation, the model was initially tested for tetragonal symmetry using the parameters specified in Table 2 and $d_{33} = 2.376 \times 10^{-9}$ m/V and $d_{31} = -0.5 d_{33}$ and $d_{15} = 0$. The results presented in Figs. 3(a, b) closely match those reported by Hwang et al.[26].

Fig. 3(a) shows the electric displacement ($D_3$) as a function of the applied electric field ($E_3$), starting from an unpoled state (random orientation). The numerical simulations were performed with 5000 crystallites, and the hysteresis loop stabilizes after completing a full cycle. The remanent polarization and strain of the ceramic, computed as the average of the crystallite spontaneous polarizations and strains, are zero at the start of the simulation. The inset labeled *A* illustrates the initial random polarization state of the crystallites, where each red dot represents the tip of a polarization vector originating from the center, forming a sphere of radius $P_0$. As the electric field increases from zero, the electric displacement gradually rises due to the dielectric effect, marking the transition from *A* to *B*. At *B*, a sharp increase in electric displacement occurs due to domain switching, leading to poling of the material. The remanent polarization nearly saturates at *C*, where the inset shows that the polarization vectors cluster near the upper pole. Upon decreasing the field from *C*, the electric displacement follows a nearly linear decline until a sudden drop occurs at *D*, corresponding to reverse domain switching (depolarization). As the field continues decreasing, the remanent polarization approaches negative saturation at *E*, where the polarization vectors cluster around the lower pole (inset *E*). When the field is increased again, a sharp jump back to positive electric displacement is observed at *F*, completing the cycle.

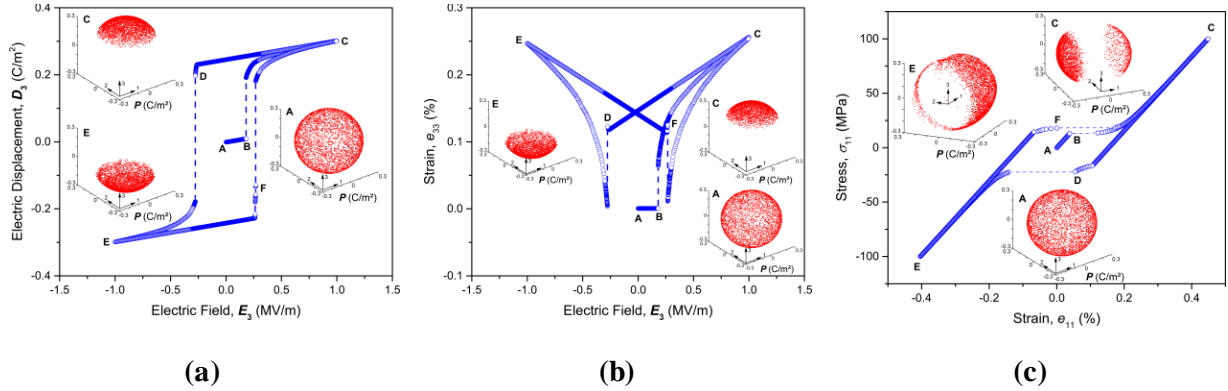

(a)    (b)    (c)

**Figure 3**. **(a)** Simulated electric displacement ($D_3$) vs. electric field ($E_3$) hysteresis loop under zero applied stress. **(b)** Simulated strain ($e_{33}$) vs. electric field ($E_3$) butterfly loop under zero applied stress. **(c)** Simulated stress ($\sigma_{11}$) vs. strain ($e_{11}$) hysteresis loop under zero applied electric field, where the loop begins with tensile stress rather than the conventional compressive stress. The material has a randomly oriented texture, and the insets depict how the polarization vectors evolve at characteristic points *A*, *B*, and *C* along the response curves.

Fig. 3(b) presents the strain ($e_{33}$) as a function of the electric field ($E_3$), corresponding to the hysteresis behavior observed in Fig. 3(a). Initially, the strain remains constant as the electric field increases from zero, indicating a nonpiezoelectric response due to the cancellation of piezoelectric contributions in the unpoled ceramic. At *B*, poling occurs, resulting in a sudden strain increase, which saturates at *C*. The strain curve follows a more gradual progression compared to electric displacement. When the electric field is reduced from *C*, the strain decreases linearly due to the piezoelectric effect, followed by a sharp drop at *D*, marking depolarization. Further reduction in the field leads to repolarization, with strain rising again to reach saturation at *E*. At this stage, the reversal of polarization alters the piezoelectric coefficients of the crystallites, leading to a sign change in the piezoelectric response. As the field increases again from *E*, the strain decreases linearly, and the cyclic process continues. The polarization vector distributions in the insets follow a similar trend to Fig. 3(a).

Fig. 3(c) depicts the average uniaxial stress ($\sigma_{11}$) versus strain ($e_{11}$) response under mechanical loading. Initially, as stress increases from zero, strain increases in accordance with linear elasticity. At *B*, 90° domain switching occurs, leading to a further increase in strain as tetragonal domains reorient, aligning their c-axes along the loading direction. At saturation point *C*, the inset shows polarization vectors aligned along the loading axis. Upon reducing the stress from *C*, the strain decreases linearly due to elastic unloading, followed by a sharp transition at *D*, where compressive stress ($\sigma_{11}$) induces domain switching, causing a jump to negative strain. Saturation is reached at *E*, where the polarization vectors align perpendicular to the compressive loading direction (inset *E*). The cycle then repeats with further unloading and reloading.

### *3.3. Polycrystalline Ferroelectric Model at the MPB Composition*

Following validation against benchmark results, we extend our simulations to explore the ferroelectric response of rhombohedral and MPB compositions under electric and mechanical loading. The MPB composition is of particular interest due to its enhanced electromechanical properties, which arise from the increased ease of polarization rotation and domain switching. In this sub-section, the electric and mechanical responses of these different phase compositions are compared under applied electric field and mechanical stress, highlighting the key differences in their polarization, strain, and stress-strain behavior.

Fig. 4(a) illustrates the polycrystalline ferroelectric model used in the simulations. The schematic represents a three-dimensional polycrystalline structure, where individual crystallites are randomly oriented. The applied electric field ($E_3$) and mechanical stresses ($\sigma_{11}$, $\sigma_{22}$, $\sigma_{33}$) are indicated, along with the associated material responses, including electric displacement ($D_3$) and strain components ($e_{11}$, $e_{22}$, $e_{33}$). The numerical simulations track the evolution of polarization and strain under external electrical and mechanical stimuli, capturing the domain switching behavior in different crystallographic phases.

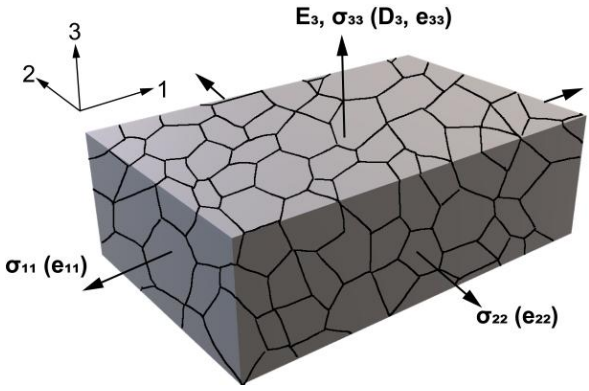
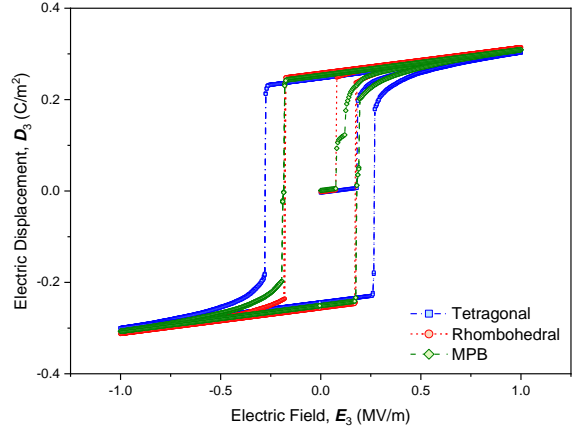
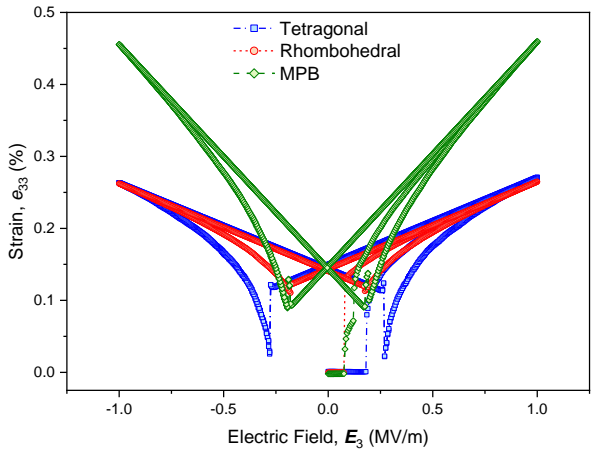
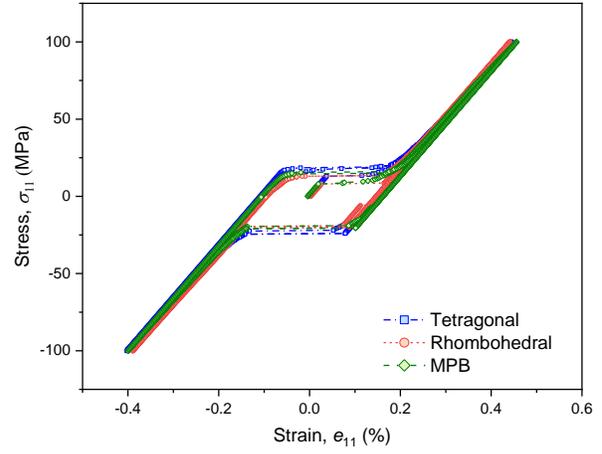

**Figure 4. (a)** Schematic representation of the polycrystalline ferroelectric model used in the simulations. **(b)** Simulated electric displacement ($D_3$) vs. electric field ($E_3$) hysteresis loops for tetragonal, rhombohedral, and morphotropic phase boundary (MPB) compositions, highlighting differences in polarization switching behavior. **(c)** Strain ($e_{33}$) vs. electric field ($E_3$) butterfly loops, showing enhanced strain response for the MPB composition due to facilitated domain reorientation. **(d)** Stress ($\sigma_{11}$) vs. strain ($e_{11}$) response under uniaxial loading for different phase compositions.

Figs. 4(b, c) present the simulated hysteresis loops for electric displacement ($D_3$) and strain ($e_{33}$), respectively, as functions of the applied electric field ($E_3$) for three different ferroelectric phase compositions: tetragonal, rhombohedral, and MPB compositions. The tetragonal phase (blue dashed line) exhibits a sharp polarization switching response, characterized by a relatively higher coercive field and distinct polarization jumps. The corresponding butterfly-shaped strain response,

similar to Fig. 3(b), shows relatively lower strain magnitudes and a smaller strain-electric field slope (piezoelectric strain coefficient) due to the limited number of available domain switching pathways. The rhombohedral phase (red dotted line) demonstrates a more gradual polarization and strain response, attributed to multi-variant domain switching and a lower coercive field. The slope of the strain-electric field curve is slightly larger than that of the tetragonal phase.

The MPB composition (green diamonds) exhibits enhanced polarization switching and the highest strain response among the three phases. This superior electromechanical performance is due to the coexistence of tetragonal and rhombohedral domains, which enables easier domain reorientation under an applied electric field. The MPB composition also has the steepest strain-electric field slope, further highlighting its advantages for piezoelectric applications. These simulation results are in excellent agreement with experimental observations reported for polycrystalline PZT for these three phase compositions (see Fig. 4 in Ref. [32]). Also, the maximization of piezoelectric strain coefficient at the MPB composition has been well reported [39].

Fig. 4(d) presents the mechanical response in terms of stress ($\sigma_{11}$) versus strain ($e_{11}$) under uniaxial loading for different phase compositions. The MPB composition exhibits the most gradual and smooth stress-strain response, attributed to the facilitated multi-variant domain reorientation. This behavior indicates that MPB materials provide an optimal balance between mechanical compliance and electromechanical coupling, reinforcing their significance in advanced actuator and sensor applications.

*3.4. Effect of Texture (or Crystallite Orientation) on Electromechanical Response*

Researchers have extensively developed textured PZT ceramics and thin films to enhance piezoelectric performance by promoting crystallite alignment along favorable crystallographic directions, as documented in the Introduction Section. Fig. 5 presents the influence of different initial crystallographic texture orientations on the piezoelectric response of polycrystalline PZT at the MPB composition under applied electric and mechanical loading. The simulations are conducted for four distinct texture cases: (001)-oriented, (100)-oriented, (111)-oriented, and randomly oriented crystallites.

For a ceramic with (001) preferred orientation, the three Euler angles ($\alpha,\beta,\gamma$) (see Fig. S1 in **Supplementary Materials**) are generated using a MATLAB random number generator to simulate crystallite alignment within the textured material. The angle $\alpha$, representing in-plane rotation, is uniformly distributed between 0 and $2\pi$. The tilt angle $\beta$ follows a normal distribution centered at zero with a standard deviation of $\pi/30$ (a variable that controls the degree of texture sharpness, where a smaller standard deviation corresponds to better texture alignment). The out-of-plane rotation $\gamma$ follows a normal distribution with a mean of 0, $\pi/2$, $\pi$, or $3\pi/2$ and a standard deviation of $\pi/30$.

For a (100) preferred orientation, the in-plane rotation $\alpha$ is uniformly distributed between 0 and $2\pi$. The tilt angle $\beta$ has a normal distribution with a mean of $\pi/2$ and a standard deviation of $\pi/30$. The rotation angle $\gamma$ is normally distributed around $\pi/2$ or $3\pi/2$ with a standard deviation of $\pi/30$.

For a (111) preferred orientation, $\alpha$ is again uniformly distributed between 0 and $2\pi$. The tilt angle $\beta$ follows a normal distribution centered at either $\cos^{-1}(1/\sqrt{3})$ or $\pi - \cos^{-1}(1/\sqrt{3})$ with a standard deviation of $\pi/30$. The out-of-plane rotation $\gamma$ follows a normal distribution with a mean of $\pi/4$, $3\pi/4$, $5\pi/4$, or $7\pi/4$ and a standard deviation of $\pi/30$.

This approach ensures that the generated Euler angles maintain the appropriate degree of texture alignment while allowing for small misorientations, simulating real-world processing conditions where perfect texture is rarely achieved. Figs. 5(a, b) illustrate the strain ($e_{33}$ and $e_{11}$) responses as a function of the applied electric field ($E_3$), for different crystallographic textures, and for a standard deviation of $\pi/10$. The (001)-oriented crystallites (black curve) exhibit the highest strain response due to favorable domain switching mechanisms. The (100)-oriented crystallites (red curve) show an intermediate strain response, followed by the response from randomly oriented crystallites (blue curve) which combines contributions from various orientations. The (111)-oriented crystallites (green curve) exhibit the lowest strain response, as polarization rotation is less favorable in this orientation. The variation in strain response confirms that crystallite orientation significantly impacts the piezoelectric properties, with the (001) texture providing the most favorable electromechanical behavior.

Experimental and theoretical studies consistently confirm that the strain response of PZT is strongly dependent on crystallographic orientation, with (001)-oriented crystallites exhibiting the highest strain response, followed by (100), random, and (111)-oriented crystallites, which show the lowest response. Phenomenological modeling (Du *et al.*, 1998) predicts that the piezoelectric coefficient $d_{33}$—and consequently the strain $e_{33}$—is maximal along [001] and decreases as the crystal tilts toward [111]. Bulk templated grain growth (TGG) experiments show that textured (001) PZT ceramics exhibit up to 2× larger strain than random polycrystals and significantly outperform (111) textures [13]. Similarly, in thin-film PZT, experimental studies [14,40] demonstrate that (001)-textured films yield 50–60% higher piezoelectric strain than (111)-textured films, attributed to enhanced polarization alignment and domain switching. Recent work on freestanding PZT films by Das *et al.* [10] confirms that strain and piezoelectric output systematically decrease from (001) to (111) orientations, with mixed textures exhibiting intermediate responses. These findings validate the model's prediction that crystallite orientation strongly influences electromechanical behavior, with (001) texture providing the most favorable strain response in both bulk and thin-film PZT.

Fig. 5(c) presents the electric displacement ($D_3$) vs. applied electric field ($E_3$) hysteresis loops for different initial crystallite orientations. The results indicate that the (001)-textured ceramic exhibits wider hysteresis loops (measured as the loop width at zero displacement) compared to the (111)-textured ceramic. Additionally, polarization reversal in the (111) texture occurs more gradually, as evidenced by the smoother transition to saturation after each polarization switching event, in contrast to sharper transitions to saturation observed in the (001) texture. This behavior suggests that domain switching dynamics are more continuous and distributed in (111)-oriented crystallites, whereas (001)-oriented crystallites undergo more abrupt polarization flips.

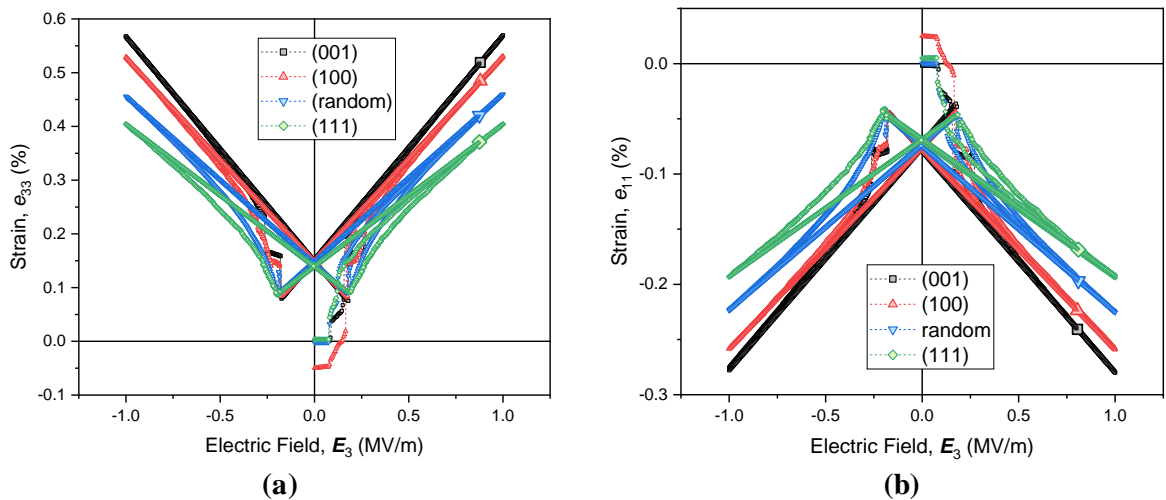

(a) (b)

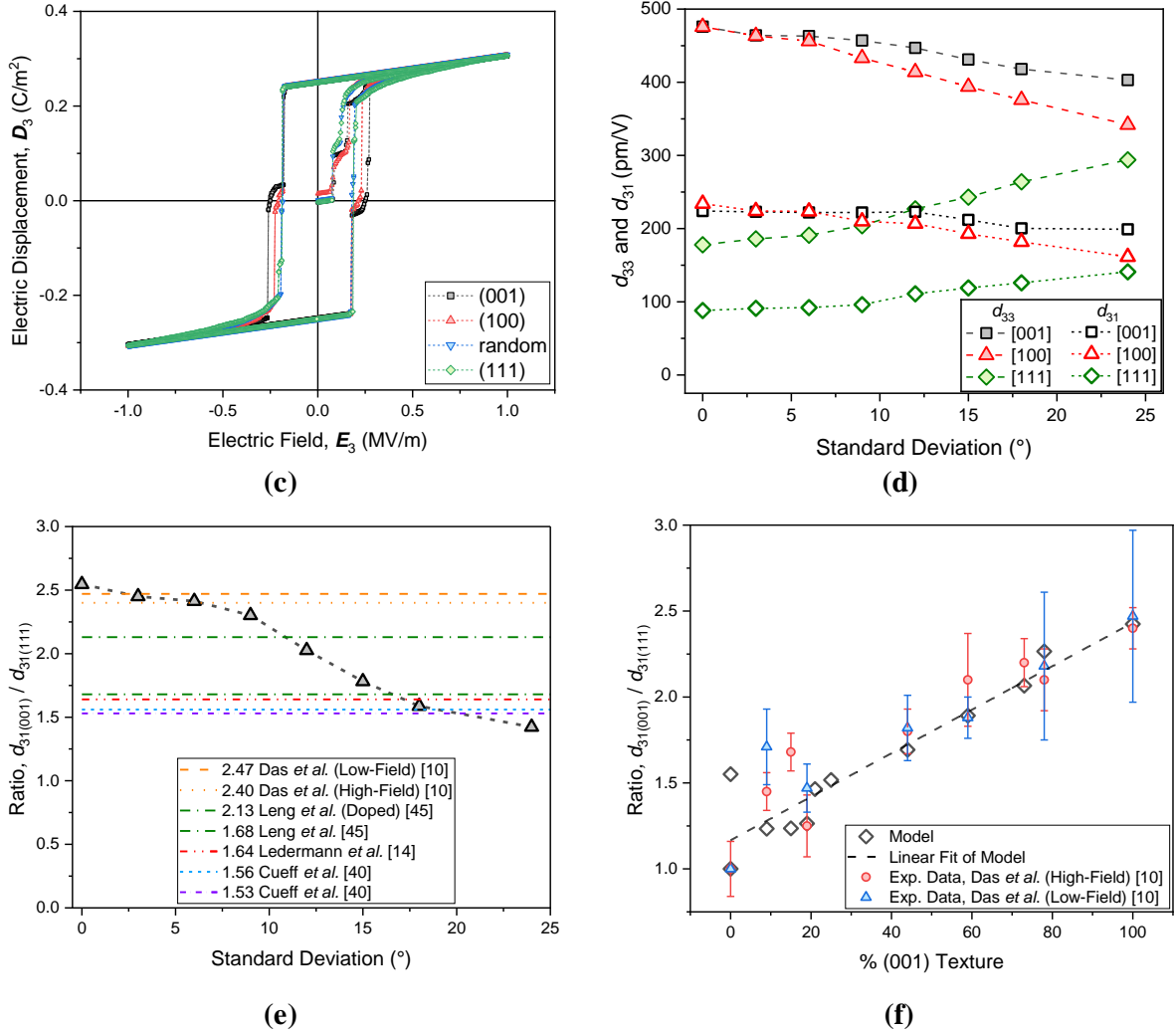

**Figure 5: (a)** Strain ($e_{33}$) vs. electric field ($E_3$) butterfly loops, illustrating that (001)-textured crystallites exhibit the highest strain response and slope ($d_{33}$), followed by (100), random, and (111) orientations. **(b)** Strain ($e_{11}$) vs. electric field ($E_3$) butterfly loops. **(c)** Electric-displacement ($D_3$) vs. $E_3$ hysteresis loops. Note that plots (a), (b), and (c) are constructed for a standard deviation of $\pi/10$. Comparison of effective piezoelectric coefficients **(d)** $d_{33}$ and $d_{31}$ for different textures, showing that (001) and (100) orientations yield the highest values and with increasing standard deviation converge towards random orientation. **(e)** Ratio of piezoelectric coefficients ($d_{31(001)}/d_{31(111)}$) compared with experimental data, demonstrating that the model predicts a maximum theoretical ratio of ~2.6, which decreases as misorientation increases, aligning well with reported experimental values. **(f)** Ratio of piezoelectric coefficients vs. % (001) texture compared with experimental data for a standard deviation of $\pi/30$.

Fig. 5(d) compares the effective piezoelectric coefficients ($d_{33}$ and $d_{31}$) for different initial crystallite orientations. Among the various textures, the (001)- and (100)-oriented crystallites exhibit the highest values of $d_{33}$ and $d_{31}$, and their responses are nearly identical when the standard deviation (or misorientation) is small. This similarity explains why many research studies do not distinguish between the (001) and (100) orientations, instead referring to them collectively as {100} or (100)/(001) texture [14]. However, as the standard deviation increases, Fig. 5(d) indicates that $d_{33}$ and $d_{31}$ for (001)-textured crystallites remain higher than those for (100)-textured crystallites, suggesting that the 001-oriented domains retain stronger piezoelectric activity even with increasing misorientation. As a result, it is better to distinguish between (001) and (100) textures. In contrast, the (111)-oriented crystallites consistently exhibit the lowest values of $d_{33}$ and $d_{31}$. Additionally, as the standard deviation increases, the values of $d_{33}$ and $d_{31}$ across all textures begin to converge, reflecting the fact that a high degree of misorientation effectively randomizes the crystallite orientations, reducing the benefits of texture alignment.

A major challenge in evaluating piezoelectric coefficients in PZT ceramics is the significant variability in reported experimental data [42,43]. This variability makes direct comparisons of absolute $d_{33}$ and $d_{31}$ values difficult. Therefore, Figure 5(e) presents a comparison of the ratio $d_{31(001)}/d_{31(111)}$ instead, allowing for a more robust comparison with experimental data. The results indicate that the maximum attainable ratio in the model is ~2.6, and as the standard deviation increases, this ratio gradually approaches → 1, consistent with increasing crystallite misalignment. Experimentally reported values also align well with the model: the highest reported ratio in literature is 2.47 [10], corresponding to freestanding PZT films with nearly 100% (001) and 100% (111) texture, where the authors estimate a standard deviation of ~5° from XRD measurements [44]. Other reported ratios fall within the range of 1.53 [40] to 1.64 [14], matching the model predictions at higher standard deviation values, indicative of less well-aligned crystallites. However, many experimental studies do not report the standard deviation of their texture measurements, a crucial parameter necessary for accurate comparison. Also, note the ratio of 2.13 and 1.68 for doped and non-doped textured PIN–PMN–PT ceramics[45], respectively.

Figure 5(f) presents a comparison between the model predictions and experimental data for the ratio $d_{31(001)}/d_{31(111)}$ as a function of (001) texture fraction (%). The experimental data, extracted from mixed-texture PZT films with varying proportions of (001), (111), and (110) orientations, show a clear trend of increasing piezoelectric coefficients with increasing (001) fraction. At the low end of the x-axis, corresponding to minimal (001) content, the data exhibit some scatter due to variability in the contributions from (111) and (110) textures—both of which dominate in this regime. Toward the high end, where the films are predominantly (001)-textured, the trend stabilizes. The model successfully captures the linear increase in the piezoelectric coefficient ratio with increasing (001) texture, demonstrating good agreement with both high-field and low-field experimental measurements reported by Das *et al.* [10].

In the next subsection, we will explore how residual stress also influences the measured $d_{31(001)}/d_{31(111)}$ ratio. However, even when accounting for residual stress, the model predicts that the maximum theoretical ratio hardly exceeds 2.6, reinforcing the limits of texture-induced enhancement in polycrystalline PZT ceramics.

Fig. 6(a) illustrates 3D spherical representations of the (001), (100), and (111) crystallographic textures, highlighting the spatial arrangement of key crystallographic planes and directions. Fig. 6(b) illustrates the evolution of polarization vectors as an electric field is applied in ferroelectric ceramics with different crystallographic textures (001), (100), and (111), for three different phase compositions: morphotropic phase boundary (MPB), tetragonal, and rhombohedral. The polarization distributions are shown at three key points in the hysteresis/butterfly loop: *A*: Initial unpoled state (distribution before applying an external field). *C*: Positive saturation (after applying a strong positive electric field). *E*: Negative saturation (after reversing the field to a strong negative value).

For (001)-Oriented Textured Crystallites: At *A*: The polarization vectors for the MPB composition are aligned along the top and bottom poles and along two rings at approximately 54.7° from the top and bottom poles (Fig. 6(a)), with some imposed standard deviation, reflecting the combination of tetragonal and rhombohedral symmetry. For tetragonal symmetry the polarization vectors cluster along the (001) directions whereas for rhombohedral symmetry the polarization vectors cluster along the (111) directions. At *C*: As the maximum positive electric field $E_3$ is applied, the polarization vectors align strongly along the field direction, leading to switching of domains to the top hemisphere. At *E*: When the field is at its maximum negative value, the polarization vectors undergo a complete reversal, flipping to the bottom hemisphere.

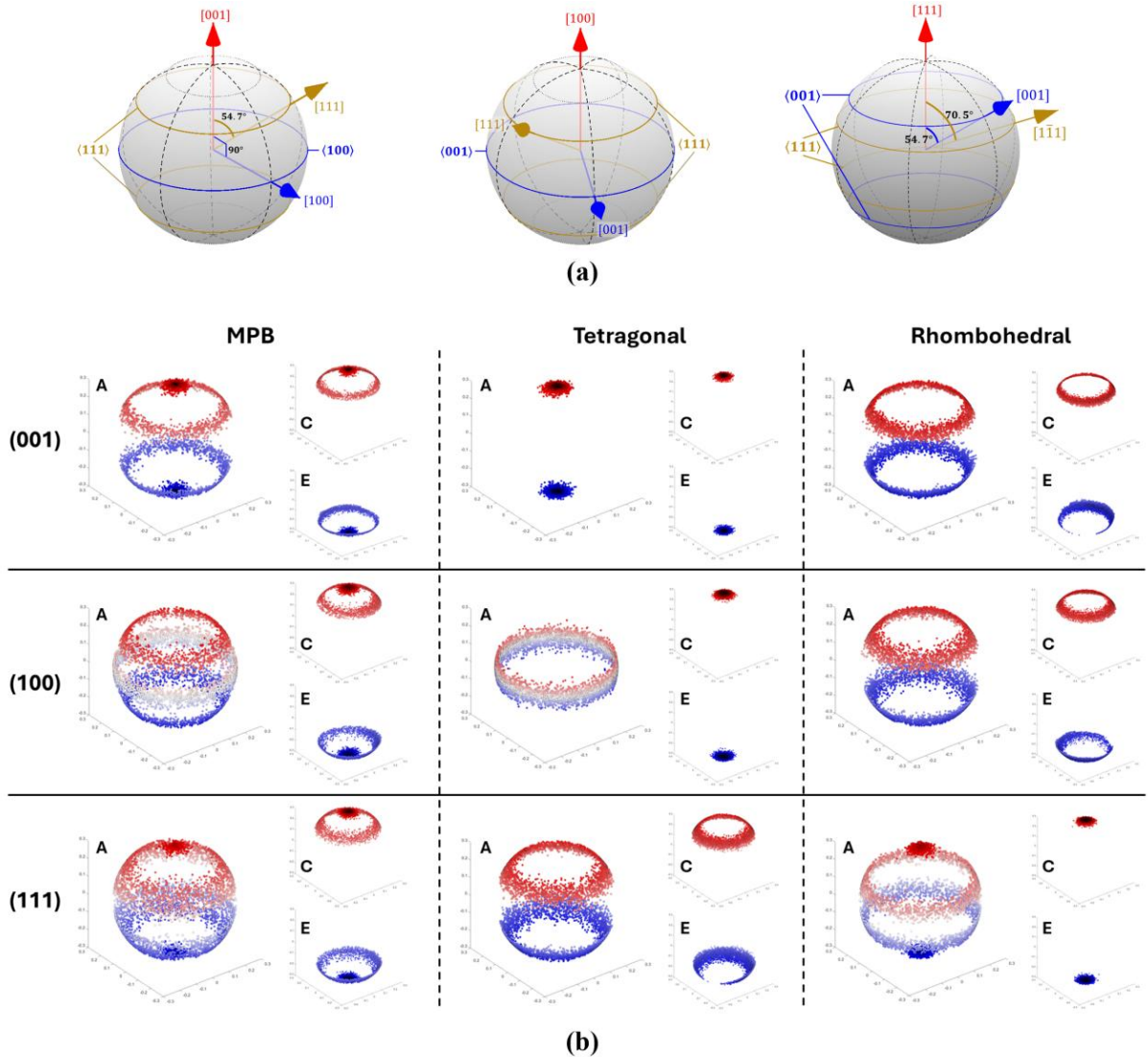

**Figure 6. (a)** 3D spherical representations of the (001), (100), and (111) crystallographic textures. Note that the angles mentioned here (54.7°, 70.5°, and 90°) are for cubic symmetry. The angles in tetragonal and rhombohedral symmetry will be slightly different from these angles because of lattice distortion. **(b)** Evolution of polarization vectors at key points in the hysteresis/butterfly loop for different crystallographic textures (001), (100), and (111) and phase compositions (MPB, tetragonal, and rhombohedral phases). The subfigures labeled *A*, *C*, and *E* represent the initial unpoled state, positive saturation, and negative saturation, respectively. The vectors are constructed for a standard deviation of π/30. The differences in domain switching behavior across textures highlight the impact of crystallographic orientation on ferroelectric properties. The color scheme follows a red-blue gradient, transitioning from white at the equator to red at the north pole and white (equator) to blue at the south pole.

<u>For (100)-Oriented Textured Crystallites</u>: At *A*: The polarization vectors for the MPB composition are aligned along the equator and along two rings at approximately 54.7° from the top and bottom poles (Fig. 6(a)). This distribution results from the combination of tetragonal and rhombohedral domain structures, where tetragonal domains align along the (001) directions on the equatorial plane, while rhombohedral domains align along the (111) directions, forming the two additional rings. At *C*: As the maximum positive electric field $E_3$ is applied, the polarization vectors for the tetragonal symmetry flip towards the field direction clustering near the top pole, whereas in the rhombohedral symmetry, the polarization vectors from the bottom ring transition toward the top ring. At *E*: When the field is at its maximum negative value, the polarization vectors from the top pole and ring have flipped to the opposite hemisphere.

For (111)-Oriented Textured Crystallites: At *A*: The polarization vectors for the MPB composition exhibit a complex distribution due to the combined influence of tetragonal and rhombohedral symmetries. For tetragonal symmetry the polarization vectors cluster along the (001) directions which in this case are along two rings at approximately 54.7° from the top and bottom poles, whereas for rhombohedral symmetry the polarization vectors cluster along the (111) directions which in this case is along the top and bottom poles and along two rings at approximately 70.5° from the poles. At *C*: As the maximum positive electric field $E_3$ is applied, the polarization vectors for the tetragonal symmetry flip towards the field direction clustering near the top ring whereas for the rhombohedral symmetry the polarization vectors from the bottom pole, bottom ring, and top ring flip to the top pole. At *E*: When the field is at its maximum negative value, the polarization vectors from the top pole and ring (state *C* in the MPB case) have flipped to the opposite hemisphere.

These differences in polarization switching behavior highlight the strong influence of texture on domain reorientation and ferroelectric response, directly impacting the piezoelectric performance of polycrystalline materials.

*3.5.    Effect of Stress (and Residual Stress) on Electromechanical Response*

Stress plays a critical role in domain switching and overall electromechanical behavior in ferroelectric and piezoelectric materials. In particular, residual stresses—which can reach several hundred MPa [12,23]—develop in sol-gel-derived lead zirconate titanate (PZT) thin films due to the constrained shrinkage of the wet film during the drying, pyrolysis, densification, and crystallization processes. These residual stresses are further influenced by thermal expansion mismatch between the film and the substrate during cooling. In most cases, biaxial tensile residual stresses develop in PZT films deposited on platinized silicon substrates (Si/SiO$_2$/Ti/Pt), leading to an in-plane clamping effect that restricts domain motion [46]. This clamping effect significantly reduces both in-plane and out-of-plane field-induced strains, thereby altering the effective piezoelectric response. The impact of residual stress is further complicated by its close relationship with film texture. Tuttle et al. [24] demonstrated that the stress state experienced by a PZT film during cooling through the Curie temperature ($T_C$) directly influences the preferred domain orientation. Tensile stresses during cooling promote an "a-domain" configuration, where the polarization vector aligns parallel to the substrate, whereas compressive stresses favor a "c-domain" configuration, where the polarization vector aligns perpendicular to the substrate. Consequently, PZT films with different textures may experience different residual stress states when deposited on stiff substrates, contributing to the variation in their linear and nonlinear ferroelectric properties [25,46]. Another key factor affecting residual stress is film thickness. Thin PZT films experience higher residual stresses compared to thicker films, leading to lower effective piezoelectric coefficients [12,46]. However, studies have shown that PZT films of different thicknesses but with similar residual stress values exhibit comparable piezoelectric responses [12]. This finding suggests that residual stress, rather than film thickness, is the dominant factor governing the electromechanical response of PZT films.

Fig. 7(a) illustrates the effect of equibiaxial tensile stress on the strain ($e_{33}$) vs. applied electric field ($E_3$) response for randomly oriented polycrystalline PZT. As tensile stress increases, the piezoelectric coefficient $d_{33}$—determined from the slope of the strain-electric field curve—decreases. This reduction in $d_{33}$ is attributed to the increased in-plane clamping effect, which restricts domain motion and reduces out-of-plane displacement, leading to a lower strain response under the same applied electric field. Similarly, Figs. 7(b, c) illustrate the decline in the piezoelectric coefficients ($d_{33}$ and $d_{31}$) with increasing tensile residual stress across different textures. However, the rates of decrease vary depending on the texture, highlighting that the impact of residual stress is highly texture dependent.

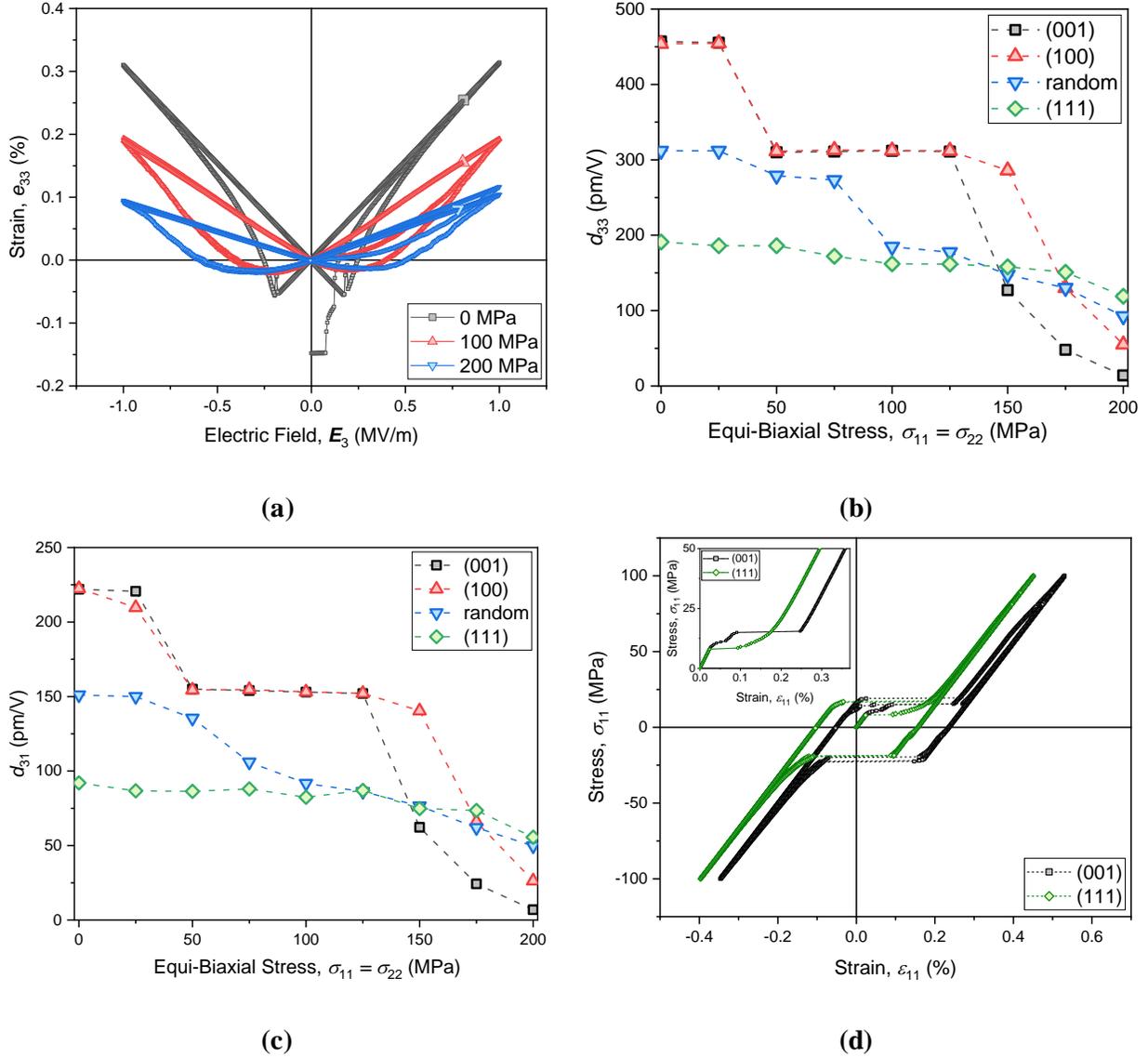

**Figure 7. (a)** Effect of equi-biaxial tensile stress on the strain ($e_{33}$) vs electric field ($E_3$) response of randomly oriented PZT. The slope, and hence the piezoelectric coefficient, decreases with increases in tensile stress. Decrease in piezoelectric coefficients **(b)** $d_{33}$, and **(c)** $d_{31}$ with increase in equi-biaxial tension and for a standard deviation of $\pi/30$. The $d_{33}$ and $d_{31}$ drops the maximum for (001) texture whereas it is relatively stable (unchanging) for the (111) texture. (d) Stress ($\sigma_{11}$) vs strain ($e_{11}$) response of (001) and (111) oriented PZT. The (001) texture shows more non-linear effects (inset) as compared to (111) texture which shows less domain-switching strains.

It is important to distinguish this observation from the previous discussion, where residual stress influenced the development of texture during film processing. In contrast, the focus here is on the electromechanical response of a material with a given texture, analyzing how residual stress alters its piezoelectric behavior rather than its initial domain structure or crystallographic alignment. As shown in Figs. 7(b, c), $d_{33}$ values exhibit the greatest drop in (001)-textured crystallites, while they remain relatively stable for (111)-textured crystallites. This suggests that (001) texture, while producing the highest piezoelectric strain response in low-stress conditions, is significantly more susceptible to degradation under high residual stress. Conversely, (111) texture demonstrates greater robustness under elevated stress conditions, maintaining more consistent electromechanical properties. At sufficiently high residual stress levels (e.g., 200 MPa in Figs. 7(b, c)), the piezoelectric response of (111)-textured crystallites or even randomly oriented crystallites may surpass that of (001)-textured crystallites, a trend that has also been observed in experimental studies [20,22]. Without accounting for residual stress, variations in piezoelectric response across different textures may be misattributed solely to crystallographic orientation, potentially leading to inaccurate comparisons and conclusions. These findings emphasize the critical importance of reporting residual stress values when comparing piezoelectric strain coefficients in PZT films. Understanding the impact of residual stress is essential for optimizing the design of PZT-based devices, particularly in applications where films are subjected to large mechanical constraints or thermal cycling.

Fig. 7(d) presents the stress ($\sigma_{11}$) vs. strain ($e_{11}$) response for (001)- and (111)-oriented PZT, illustrating the tension-unloading-compression-unloading cycle. While ceramics are typically designed to operate under compressive loading due to their inherent flaw sensitivity in tension, thin PZT films in $d_{31}$-flexural mode experience both tensile and compressive stresses. Therefore, the tension-compression response is analyzed here. The inset highlights a key distinction in ferroelastic behavior: (001)-textured PZT exhibits greater non-linearity, characterized by a larger strain jump during domain switching, whereas (111)-textured PZT shows reduced domain-switching strains and less deviation from the initial elastic path. This indicates that (001)-oriented crystallites undergo more pronounced ferroelastic switching, leading to greater hysteresis, whereas (111)-oriented crystallites exhibit a more stable and linear stress-strain response. The increased ferroelastic non-linearity in (001)-textured PZT compared to (111)-textured PZT has been previously reported in studies on MPB compositions [9], further reinforcing the texture-dependent nature of ferroelastic switching in PZT.

Fig. 8 illustrates the evolution of polarization vectors under the application of a maximum electric field ($E_3$) in ferroelectric ceramics with different crystallographic textures—(001), (100), and (111)—at the MPB composition for four different levels of equi-biaxial stress. The results highlight how increasing stress alters domain configurations and impacts the effective piezoelectric coefficients ($d_{33}$ and $d_{31}$).

(001)-Textured Crystallites: In the absence of applied stress and electric field, the polarization vectors in (001)-textured crystallites exhibit a characteristic distribution reflecting the coexistence of tetragonal and rhombohedral domains. When a 25 MPa equi-biaxial stress is applied, tetragonal domains shift toward the equator, while rhombohedral domains remain near their original positions. Under the subsequent application of a maximum positive electric field, the tetragonal domains realign near the north pole, while rhombohedral domains cluster along the 54.7° ring in the northern hemisphere. However, as the in-plane stress increases to 50 MPa, the electric field is no longer sufficient to induce complete 90° domain switching in tetragonal crystallites. Instead, the polarization vectors form two distinct rings: one at the equator (tetragonal domains) and another at ~54.7° (rhombohedral domains). This reduced domain switching results in a drop in $d_{33}$ from 455 to 311 pC/N. For stress levels up to 125 MPa, the polarization configuration remains largely unchanged, and the slope of the butterfly loop continues to decrease gradually. However, at 150 MPa, the large in-plane stress induces a significant interphase domain switching, with a large fraction of rhombohedral domains transforming into tetragonal domains near the equator. As a result, $d_{33}$ drops sharply from 311 to 127 pC/N, with 79% of domains becoming tetragonal under the applied maximum electric field. Beyond this point, the slope strictly decreases, and at 200 MPa, $d_{33}$ is reduced to just 14 pC/N. Both $d_{33}$ and $d_{31}$ follow the same overall pattern, decreasing progressively with increasing stress, Figs. 7(b, c).

(100)-Textured crystallites: The (100)-textured crystallites exhibit a response similar to (001)-textured crystallites under increasing equi-biaxial stress. However, beyond 150 MPa, the numerical values of $d_{33}$ and $d_{31}$ begin to diverge slightly between the two textures, despite following the same overall trend. This discrepancy arises because, even with the same standard deviation, the polarization vectors in (100)-textured crystallites are more widely dispersed compared to those in (001)-textured crystallites. The increased dispersion is attributed to differences in angular constraint and distribution geometry. In (001)-textured crystallites, polarization vectors are clustered near the poles, resulting in a more localized spread. In contrast, (100)-textured crystallites distribute their polarization vectors along the equatorial plane, covering a broader angular region. Additionally, stereographic projection effects further amplify this apparent spread. This greater dispersion leads to a lower degree of interphase domain switching (rhombohedral → tetragonal) under the same applied stress. As a result, at 150 MPa, only 55% of the domains in the (100) texture undergo tetragonal transformation and accumulate near the equator, compared to 79% in the (001) texture. This reduced transformation results in higher $d_{33}$ and $d_{31}$ values in (100)-textured crystallites relative to (001)-textured crystallites at equivalent stress levels.

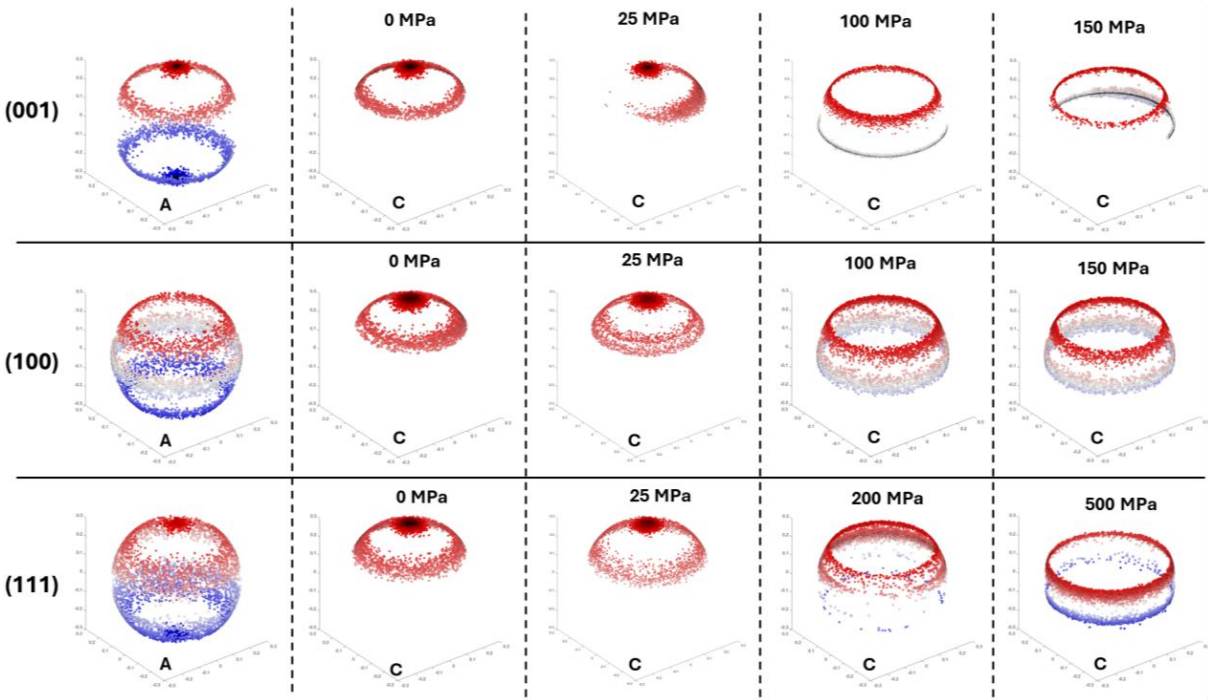

**Figure 8.** Polarization vectors in textured-MPB PZT at point *A* (initial configuration) and their evolution under applied in-plane equi-biaxial tension. Snapshots are shown for point *C* (1 MV/m positive electric field) at progressively increasing tensile stress levels. The vectors are constructed for a standard deviation of π/30.

<u>(111)-Textured crystallites</u>: Upon applying an electric field, the polarization vectors from the lower hemisphere shift to the upper hemisphere (see Fig. 6). However, due to the absence of possible domain switching near the equator, even at high in-plane stress levels (up to 175 MPa), there is minimal change in the butterfly loop slope. Unlike (001) and (100) textures, (111)-textured crystallites experience less interphase domain switching, and the overall piezoelectric response remains relatively stable. The primary effect of stress is a minor rearrangement of domain orientations, leading to only a slight reduction in $d_{33}$ and $d_{31}$ rather than a significant degradation. At 200 MPa, a small but noticeable decrease in both $d_{33}$ and $d_{31}$ occurs, primarily due to interphase domain switching from tetragonal to rhombohedral domains. Since rhombohedral domains are closer to the equator in (111)-textured crystallites, the stress-driven transformation follows an opposite trend compared to (001) and (100) textures. However, if stress levels reach 500 MPa, even under a maximum electric field, two distinct polarization rings form in the northern and southern hemispheres, effectively neutralizing the piezoelectric response and reducing butterfly loop slopes to near zero.

Randomly Oriented crystallites: For randomly oriented crystallites, the response exhibits a continuous decrease in piezoelectric coefficients with increasing stress. Unlike the more distinct transitions observed in highly textured samples, interphase domain switching begins gradually between 125 MPa and 150 MPa, leading to a progressive degradation of piezoelectric performance.

The interplay between equibiaxial stress, texture, and domain switching mechanisms significantly impacts the electromechanical response of PZT at MPB. While (001)- and (100)-textured crystallites exhibit superior piezoelectric performance in low-stress conditions, their response degrades significantly beyond 150 MPa due to extensive stress-induced domain transformations. In contrast, (111)-textured crystallites maintain a more stable electromechanical response due to their reduced susceptibility to stress-induced domain switching. These findings emphasize the importance of considering stress (and residual stress) effects when evaluating piezoelectric performance, particularly in thin-film applications where large in-plane stresses are prevalent.

## 3.6. *Fitting Experimental Data*

In this final section, we demonstrate that the model can effectively fit experimental data for both randomly oriented and textured PZT. Fig. 9 presents the comparison between numerical simulations and experimental measurements, with the parameters used for fitting provided in Table 3.

For randomly oriented PZT at the MPB composition, Fig. 9(a) shows that the model successfully captures the $e_{33}$ vs. $E_3$ butterfly loop from the experimental work of Saito *et al.*[47]. However, at high electric fields, the experimental data deviates from the numerical prediction, especially on the side of the positive electric field. This discrepancy arises because the model, like most existing models, does not fully account for the reduction in extrinsic contributions—particularly domain wall motion saturation—which leads to an overestimation of strain in the high-field regime.

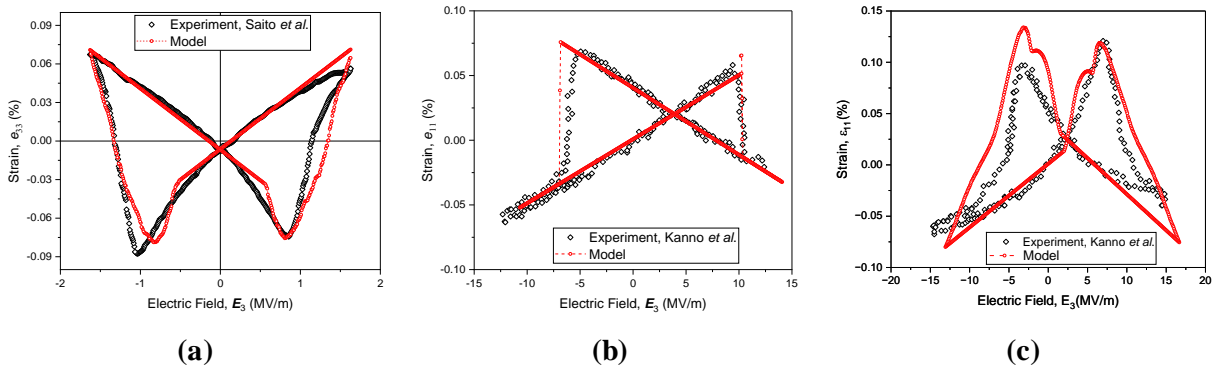

**Figure 9**. Comparison between experimental data and computational model predictions for: **(a)** randomly oriented MPB PZT[47], **(b)** (001)-textured PZT thin film [41], and **(c)** dominantly (111)-textured PZT thin film [41]. To facilitate direct comparison, the model curves have been translated and shifted to align with the experimental intersection points while preserving the relative trends in electromechanical response.

To validate the model for textured PZT, we referenced experimental cantilever deflection data from the work of Kanno et al.[41], due to the limited availability of textured bulk PZT data in the literature. The authors fabricated $d_{31}$-type PZT thin films (2.9 µm and 2.5 µm thick) on two different substrates: (100)Pt/(100)MgO, which resulted in a predominantly (001)-textured PZT, and (111)Pt/Ti/Si, which produced a predominantly (111)-textured PZT. Cantilever tip deflection was measured using a bipolar sine wave signal (±35V, 10 Hz), and the recorded deflection vs. voltage data was converted to strain ($e_{11}$) vs. electric field ($E_3$) using the multimorph deflection model by DeVoe et al. [48]. This model requires input parameters such as flexural rigidities and layer thicknesses, which allow for the calculation of $d_{31}$ as a function of $E_3$. The strain reported here is the $e_{11}$ strain, which is the product of $d_{31}$ and $E_3$. The film in-plane Young's modulus of Pt was taken from the work of Das et al. [49], while the thin Au layer was assumed to have a modulus of 40 GPa.

Fig. 9(b) shows the $e_{11}$ strain response for the cantilever with a c-axis oriented PZT film on MgO. The experimental strain vs field curve exhibits an ideal butterfly loop with distinct 180° domain switching. Apart from the 180° switching, the strain in the PZT layer shows excellent linearity, attributed to the highly oriented c-axis texture of the PZT film. The model successfully captures both domain switching behavior and the linearity of the response. In contrast, the strain-field response for the PZT film on a Si substrate exhibits significant displacement hysteresis, as shown in Fig. 9(c). This behavior arises because, in addition to intrinsic lattice motion, the piezoelectric strain is also influenced by off-axis domain reorientation, particularly 90° domain switching. The theoretical model effectively captures this effect, demonstrating its capability to account for both lattice-driven and domain-driven strain contributions.

**Table 3**: Parameters used for fitting experimental data

| Parameter | Saito *et al.*[47] | Kanno *et al.* [001][41] | Kanno *et al.* [111][41] |
|---|---|---|---|
| $\bar{E}_0^{90}$ (MV/m) | 0.7 | 4.2 | 2 |
| $\bar{E}_0^{180}$ (MV/m) | 2 | 12 | 12 |
| $\bar{E}_0^{71\_109}$ (MV/m) | 0.7 | 5 | 4 |
| $\bar{E}_0^{55\_125}$ (MV/m) | 1.5 | 10 | 10 |
| $d_{ijk}$ | $d_{ijk}/5$ (tetragonal) | $d_{ijk}/40$ (tetragonal) | $d_{ijk}/25$ (tetragonal) |
| $N$ | 500 | 2000 | 5000 |
| $e_{33}$ shift | 0.05% | - | - |
| $e_{11}$ shift | - | 0.02% | 0 |
| $E_3$ shift (MV/m) | - | 1.8 ($E_3$ axis flipped) | 1.7 ($E_3$ axis flipped) |
| Standard Deviation (rad) | N/A (random texture) | 0 | $\pi/10$ |
| Biaxial Stress (MPa) | 45 | 0 | 200 (uniaxial) |

The numerical model, thus, successfully reproduces experimental butterfly loops for both randomly oriented and textured PZT. The ability of the model to capture domain switching mechanisms and texture-dependent strain behavior makes it a valuable tool for understanding and optimizing electromechanical responses in PZT-based devices.

## 4. Conclusion

This computational study demonstrates that crystallographic texture, along with both applied and residual stress, plays a critical role in governing the electromechanical response of PZT near the morphotropic phase boundary (MPB). The results confirm that piezoelectric performance is maximized near the MPB, where coexisting tetragonal and rhombohedral phases enable enhanced domain switching. Among different orientations, (001)-textured ceramics exhibit the highest piezoelectric coefficients under low stress conditions, with this enhancement persisting even when the standard deviation in crystallite orientation reaches ~25°, underscoring the resilience of partial alignment. The model further distinguishes between (001)- and (100)-textured configurations in the presence of orientation variability. The ratio $d_{31(001)}/d_{31(111)}$ is found to increase nearly linearly with the (001) texture fraction, in close agreement with experimental trends, and reaches a theoretical maximum of ~2.6, consistent with experimental values ranging from 1.5 – 2.5. These findings highlight the importance of not only identifying the dominant orientation but also quantifying the texture distribution when evaluating or designing ferroelectric ceramics. Stress-induced changes in domain switching dynamics are also shown to be pivotal. While (001)- and (100)-textured materials offer superior performance at low stress, their piezoelectric response degrades significantly beyond ~150 MPa due to ferroelastic switching. In contrast, (111)-textured or randomly oriented crystallites exhibit a more stable response, aligning well with reported experimental behavior. This distinction is especially relevant for thin-film applications, where high in-plane residual stresses can significantly influence electromechanical performance. Together, these insights offer a predictive framework for understanding and tailoring piezoelectric behavior in MPB PZT. By capturing the coupled effects of texture, stress, and phase switching, this work provides valuable guidance for the design and optimization of next-generation ferroelectric materials and devices.

## 5.     Acknowlegement

The authors acknowledge support from the Indian Institute of Science through seed grants (IE/CARE-21-0355) and (IE/RERE-21-0610).

## 6.     Data and Code Availability

The code is available on Figshare (https://doi.org/10.6084/m9.figshare.28600262), and the access link is also provided in the Supplementary Materials.

## 7.     Conflict of Interest Statement

The authors declare no competing financial or non-financial interests.